\documentclass[reprint,superscriptaddress,aps,twocolumn,showpacs]{revtex4-1}
\usepackage{amssymb}
\usepackage{amsmath}
\usepackage{graphicx}
\usepackage{multirow}
\usepackage{hyperref}
\usepackage{color}

\allowdisplaybreaks

\begin{document}
\title{Study of light tetraquark spectroscopy}

\author{Zheng Zhao}
\email[]{zhaozheng1022@hotmail.com}
\author{Kai Xu}
\author{Attaphon Kaewsnod}
\affiliation{School of Physics and Center of Excellence in High Energy Physics and Astrophysics, Suranaree University of Technology, Nakhon Ratchasima 30000, Thailand}
\author{Xuyang Liu}
\affiliation{School of Physics and Center of Excellence in High Energy Physics and Astrophysics, Suranaree University of Technology, Nakhon Ratchasima 30000, Thailand}
\affiliation{School of Physics, Liaoning University, Shenyang 110036, China}
\author{Ayut Limphirat}
\author{Yupeng Yan}
\email[]{yupeng@sut.ac.th}
\affiliation{School of Physics and Center of Excellence in High Energy Physics and Astrophysics, Suranaree University of Technology, Nakhon Ratchasima 30000, Thailand}

\date{\today}

\begin{abstract}
\indent We calculate the masses of the $qq\bar q\bar q$ tetraquark ground state and first radial excited state in a constituent quark model where the Cornell-like potential and one-gluon exchange spin-spin coupling are employed. The three coupling parameters for the Cornell-like potential and one-gluon exchange spin-spin coupling are proposed mass-dependent in accordance with Lattice QCD data, and all model parameters are predetermined by studying light, charmed and bottom mesons.
The theoretical predictions for light tetraquarks are compared with the observed exotic meson states in the light-unflavored meson sector, and tentative assignments are suggested. The work suggests  that the $f_0(1500)$ and $f_0(1710)$ might be ground light tetraquark states with $J=0$.
\end{abstract}

\maketitle

\section{Introduction}\label{sec:Int}

Low-lying baryons and mesons, even for the light quark sector (with the exception of the would-be Goldstone bosons of the chiral symmetry breaking, firstly the pions), can be reasonably described in non-relativistic quark models (NQM), where the interaction of constituent quarks is interpreted in terms of potentials which are usually phenomenologically motivated~\cite{Brambilla:2019esw}.

\begin{figure}[tb]
\label{status}
\centering
\includegraphics[width=0.48\textwidth]{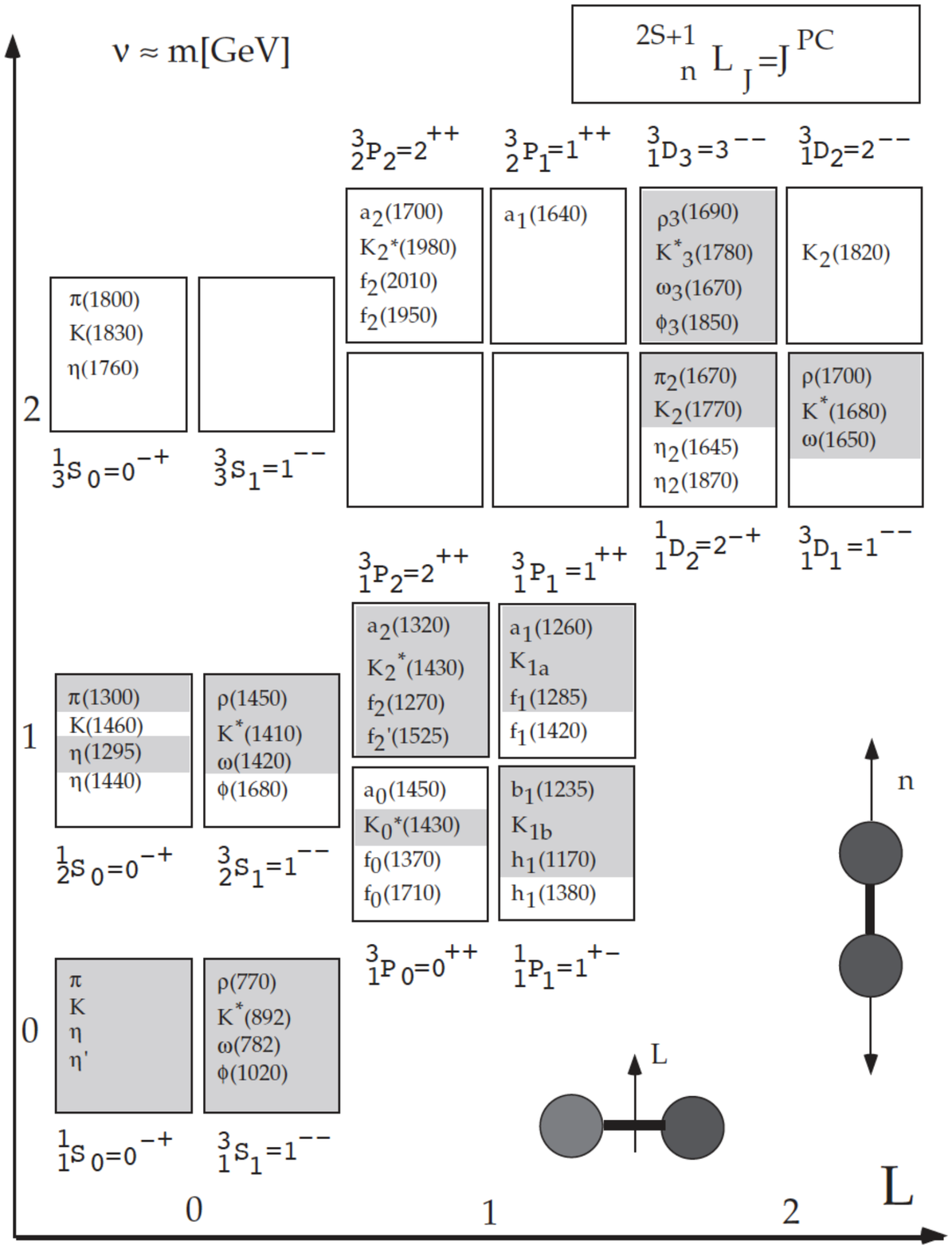}
\caption{Tentative $q\bar q$ mass spectrum for the three light quarks in $SU(3)$ symmetry. The shaded assignments are clear and definitive~\cite{Amsler:2004ps}.}
\end{figure}

Theoretical predictions for meson mass spectrum can be found in \cite{Godfrey:1985xj, Vijande:2004he, Ebert:2009ub, Xiao:2019qhl, Li:2020xzs}. Most $L=0$ and $1$ meson nonets in NQM can be easily associated with the well established experimental candidates, with reasonable flavor symmetry breaking and binding assumptions, except for the scalar $^3P_0$ nonet for which there are too many observed candidates~\cite{Hagiwara:2002fs}. The experimental status of light mesons is shown in Figure~\ref{status} in an approximated mass scale, where the mesons of an isovector, a strange isodoublet and two isoscalars are grouped together to represent a flavor nonet. The total angular momentum $J$ , orbital excitation $L$, spin multiplicity $2S +1$, and radial excitation $n$ of the states are used for classification. The vertical scale is $v=n+L-1$, and the horizontal scale is the $L$. The ground state pseudoscalars ($J^{PC} = 0^{-+}$) and vectors ($1^{--}$) are well established. However, a number of predicted radial excitations ($n>1$) and orbital excitations ($L>0$) are still missing and some observed meson candidates do not fit into quark model conventional meson mass spectra easily~\cite{Amsler:2004ps}.

The researches on conventional $q\bar q$ states conducted for more than a half century provide us with a good knowledge to understand their underlying structures. Exotic meson states such as  glueballs, hybrids, and tetraquarks have been widely studied in the past two decades, especially focusing on the states having same quantum numbers as conventional $q\bar q$ systems. We will briefly review and discuss some typical exotic mesons in Sec.~\ref{sec:DIS}.

\indent In this work, we apply the same model as the one in Ref.~\cite{Zhao:2021bma} to predict the mass of all possible light tetraquark configurations.
The paper is organized as follows. In Sec.~\ref{sec:TM}, we work out the possible configurations of color, spin, and spatial degrees of freedom of tetraquark states. In Sec.~\ref{sec:DIS}, tetraquark mass spectra are evaluated in the constituent quark model applied in previous works~\cite{Xu:2019fjt, Xu:2020ppr, Zhao:2021bma, Zhao:2021cdv}, and the theoretical results are compared with experimental data and  tentative assignments for light tetraquarks are suggested. A summary is given in Sec.~\ref{sec:SUM}.

\section{\label{sec:TM}THEORETICAL MODEL}

{The quark and antiquark transform under the fundamental and conjugate representations of {$SU(n)$}, respectively, with {$n=3, 2, 2$} for the color, spin, and flavor degree of freedom.}
The color part of the wave function of a tetraquark is a $[222]_1$ singlet of the $SU_c(3)$ group.
The Young tabloids $[2]_6$ and $[11]_{\bar 3}$, and $[22]_{\bar 6}$ and $[211]_3$ of the the $SU_c(3)$ group characterize the permutation symmetry of the two quarks cluster ($qq$), and the two antiquarks cluster ($\bar q\bar q$) of tetraquark states, respectively. Thus the $[222]_1$ color singlet of tetraquark states demands the following configurations
\begin{eqnarray}
\psi^c_{6\otimes\bar 6} & \equiv &  [2]_6(q_1q_2)\otimes[22]_{\bar 6}(\bar q_3\bar q_4), \nonumber \\
\psi^c_{\bar 3\otimes3} & \equiv & [11]_{\bar 3}(q_1q_2)\otimes[211]_3(\bar q_3\bar q_4).
\end{eqnarray}

\indent Considering that a $qq\bar q\bar q$ tetraquark state of four light quarks must be a color singlet and antisymmetric simultaneously under any permutation between identical quarks, one gets all the possible color-spatial-spin-flavor configurations of the $qq$ and $\bar q\bar q$ cluster listed in Table~\ref{tab:ssf}.
{Here we have used that the fundamental and conjugate representations of {$SU_s(2)$} and {$SU_f(2)$} for quarks and antiquarks are the same.}

\begin{table}[tb]
\caption{All possible color-spatial-spin-flavor configurations of $qq$ and $\bar q\bar q$ cluster.}
\label{tab:ssf}
\begin{ruledtabular}
\begin{tabular}{ccc}
\multirow{4}*{$qq$} & \multirow{2}*{$\psi^c_{[2]}\psi^{osf}_{[11]}$} & $\psi^c_{[2]}\psi^o_{[2]}\psi^s_{[11]}\psi^f_{[2]}$
\\
&&$\psi^c_{[2]}\psi^o_{[2]}\psi^s_{[2]}\psi^f_{[11]}$
\\ \cline{2-3}
&\multirow{2}*{$\psi^c_{[11]}\psi^{osf}_{[2]}$} & $\psi^c_{[11]}\psi^o_{[2]}\psi^s_{[11]}\psi^f_{[11]}$
\\
&& $\psi^c_{[11]}\psi^o_{[2]}\psi^s_{[2]}\psi^f_{[2]}$
\\
\hline
\multirow{4}*{$\bar q\bar q$} & \multirow{2}*{$\psi^c_{[22]}\psi^{osf}_{[11]}$} & $\psi^c_{[22]}\psi^o_{[2]}\psi^s_{[11]}\psi^f_{[2]}$
\\
&&$\psi^c_{[22]}\psi^o_{[2]}\psi^s_{[2]}\psi^f_{[11]}$
\\ \cline{2-3}
&\multirow{2}*{$\psi^c_{[211]}\psi^{osf}_{[2]}$} & $\psi^c_{[211]}\psi^o_{[2]}\psi^s_{[11]}\psi^f_{[11]}$
\\
&& $\psi^c_{[211]}\psi^o_{[2]}\psi^s_{[2]}\psi^f_{[2]}$
\end{tabular}
\end{ruledtabular}
\end{table}

The possible spin combinations are
\begin{eqnarray}
\psi^{S=0,1,2}_{(1\otimes1)} & \equiv & \left[\psi_{[s=1]}^{qq}\otimes\psi_{[s=1]}^{\bar q \bar q}\right]_{S=0,1,2}, \nonumber\\
\psi^{S=1}_{(1\otimes0)} & \equiv & \psi_{[s=1]}^{qq}\otimes\psi_{[s=0]}^{\bar q \bar q}, \nonumber \\
\psi^{S=0}_{(0\otimes0)} & \equiv & \psi_{[s=0]}^{qq}\otimes\psi_{[s=0]}^{\bar q \bar q}.
\end{eqnarray}
The color and spin wave functions take the same forms as the ones for charmonium-like tetraquarks in Ref.~\cite{Zhao:2021bma}.

\indent The relative Jacobi coordinates and the corresponding momenta of light tetraqaurk are defined as
\begin{flalign}\label{eqn::ham}
&\vec x_1=\frac{1}{\sqrt 2}(\vec r_1-\vec r_3), \nonumber \\
&\vec x_2=\frac{1}{\sqrt 2}(\vec r_2-\vec r_4), \nonumber \\
&\vec x_3=\frac12({\vec r_1+\vec r_3}-{\vec r_2-\vec r_4}), \nonumber \\
&\vec x_0=\frac14({\vec r_1+\vec r_2+\vec r_3+\vec r_4}), \nonumber \\
&\vec{p}_{i}= u_i \frac{d\vec x_i}{dt},
\end{flalign}
where $u_i$ are the reduced quark masses defined as
\begin{flalign}\label{eqn::rqm}
u_1=u_2=u_3=m_{u,d},
\end{flalign}
where $\vec{r}_{j}$ and $m_{u,d}$ are the coordinate and mass of the $u$ and $d$ quark.

\indent The total tetraquark spatial wave function may be expanded in the complete basis formed by the functions,
\begin{eqnarray}\label{eqn::spatial}
\psi_{NL} &=&
\sum_{\{n_i,l_i\}}
 A(n_{1},n_{2},n_{3},l_{1},l_{2},l_{3}) \nonumber \\
&& \times \psi_{n_{1}l_{1}}(\vec x_1\,) \otimes\psi_{n_{2}l_{2}}(\vec x_2\,)\otimes\psi_{n_{3}l_{3}}(\vec x_3\,)
\end{eqnarray}
where $\psi_{n_{i}l_{i}}$ are harmonic oscillator wave functions and the sum $\{n_i,l_i\}$ is over $n_{1}, n_{2}, n_{3}, l_{1}, l_{2}, l_{3}$. $N$ and $L$ are the total principle quantum number and orbital angular momentum number of the $qq\bar q\bar q$ tetraquark respectively. One has $N= (2n_{1}+ l_{1})+(2n_{2}+ l_{2})+(2n_{3}+l_{3})$.
The spatial wave functions $\psi_{NL}$ are employed as complete bases to study tetraquark states with other interactions. The bases size is N=14 in the calculations, and the length parameter of harmonic oscillator wave functions is adjusted to 450MeV to get the best eigenvalue. The complete bases of the tetraquarks are imported from Ref.~\cite{Zhao:2021bma}.

\indent The nonrelativistic Hamiltonian for studying the meson and tetraquark systems, which is the same as the one in Ref.~\cite{Zhao:2021bma}, takes the form,
\begin{flalign}\label{eqn::ham}
H = &H_0+ H_{hyp}^{OGE}, \nonumber \\
H_{0} = &\sum_{k=1}^{N} (\frac12M^{ave}_{k}+\frac{p_k^2}{2m_{k}}), \nonumber \\
&+\sum_{i<j}^{N}(-\frac{3}{16}\lambda^{C}_{i}\cdot\lambda^{C}_{j})(A_{ij} r_{ij}-\frac{B_{ij}}{r_{ij}}),  \nonumber \\
H_{hyp} = &\sum_{i<j}C_{ij}{\lambda^{C}_{i}\cdot\lambda^{C}_{j}}\,\vec\sigma_{i}\cdot\vec\sigma_{j}.
\end{flalign}
where $m_k$ are the constituent quark masses, and $M^{ave}_k$ is the spin-averaged mass. $\lambda^C_{i}$ and $\vec\sigma_i$ in Eq. (\ref{eqn::ham}) are the quark color operator and the spin operator respectively.

\indent The string tension coefficient $A$ and Coulomb coefficient $B$ of Cornell potential $V(r)=Ar-B/r $ may take different values for different hadron sectors while being fitted to experimental data, which happens not only in quark model studies but also in lattice QCD studies~\cite{Kawanai:2011xb, Ikeda:2011bs}. The Cornell potential is employed to fit the QCD data of interquark potentials ($V_{q\bar q}$) at finite quark mass ($m_{q}$) in Refs.~\cite{Kawanai:2011xb, Ikeda:2011bs}, and the fitting results suggest that both the string tension coefficient $A$ and Coulomb coefficient $B$ are mass dependent,  taking the forms $A=a+b\ {m_q}$ and $B=B_0\sqrt{1/{m_{q}}}$, respectively. A mass dependent Coulomb coefficient $B$ is also suggested in Ref.~\cite{Bali:2000gf}. For more detailed discussion, we refer to Ref.~\cite{Zhao:2021bma}.
The hyperfine coefficient $C_{ij}$ is proposed to have the same mass dependence as the coulomb-like interaction, assuming that the hyperfine interaction and Coulomb-like interaction are from the same route of one gluon exchange.

\indent Therefore, $A_{ij}$, $B_{ij}$, and $C_{ij}$ in Eq. (\ref{eqn::ham}) are proposed to be mass dependent coupling parameters, taking the form
\begin{eqnarray}
A_{ij}= a+bm_{ij},\;B_{ij}=B_0 \sqrt{\frac{1}{m_{ij}}},\;C_{ij} =C_0 \sqrt{\frac{1}{m_{ij}}},
\end{eqnarray}
with $a$, $b$, $B_0$, and $C_{0}$ being constants. The reduced mass of $i$th and $j$th quarks, $m_{ij}$, defined as $\;m_{ij}=\frac{2 m_i m_j}{m_i+m_j}$.
The four constituent quark masses and four model coupling parameters are determined by comparing the theoretical and experimental masses of conventional mesons as follows,
\begin{eqnarray}\label{eq:nmo1}
&m_{u,d} = 380 \ {\rm MeV}\,, \quad m_s = 550 \ {\rm MeV}\,, \nonumber\\
&m_c = 1270 \ {\rm MeV}\,, \quad m_b = 4180 \ {\rm MeV}\,,  \nonumber\\
&a=67413 \ {\rm MeV^2}, \quad b=35 \ {\rm MeV}\,,  \nonumber\\
&B_0=31.7 \ {\rm MeV^{1/2}}\,, \quad C_0=-188.8 \ {\rm MeV^{3/2}}\,.
\end{eqnarray}

\indent We employ $m_{u,d} = 380 \ {\rm MeV}$ in the work, which is slightly larger than the conventional value around $350$ MeV. It is found that a bigger $m_{u,d}$ ranging from 380 to 450 MeV leads to good fitting results to the mass of $\rho(770)$, $\rho(1450)$, $D^0(1870)$, $D^*(2010)^0$, $B^0(5279)$, and $B^*(5325)$, and when $m_{u,d} = 380 \ {\rm MeV}$ the fitting results are $788$, $1455$, $1876$, $2043$, $5218$, and $5371$ MeV respectively. The other fitting results and details are shown in Ref.~\cite{Zhao:2021bma}.

\section{\label{sec:DIS}Results and discussion}

We evaluate the mass spectra of the ground and first radial excited light tetraquarks in the Hamiltonian in Eq. (\ref{eqn::ham}) including the color-spin interaction $H_{hyp}$ which may mix up different color-spin configurations. There is no mixing between different flavor configurations as we treat the Hamiltonian flavor independent. Because of the cross terms,
\begin{flalign}
&\langle\psi^c_{\bar 3\otimes3}\psi^{S=0}_{(0\otimes0)}|\vec \lambda_i\cdot\vec \lambda_j\,\vec \sigma_i\cdot\vec \sigma_j|\psi^c_{6\otimes\bar 6}\psi^{S=0}_{(1\otimes1)}\rangle=8\sqrt 6, \nonumber \\
&\langle\psi^c_{\bar 3\otimes3}\psi^{S=0}_{(1\otimes1)}|\vec \lambda_i\cdot\vec \lambda_j\,\vec \sigma_i\cdot\vec \sigma_j|\psi^c_{6\otimes\bar 6}\psi^{S=0}_{(0\otimes0)}\rangle=8\sqrt 6,
\end{flalign}
eigenstates of the Hamiltonian are linear combinations of $\psi^c_{\bar 3\otimes3}\psi^{S=0}_{(0\otimes0)}$ and $\psi^c_{6\otimes\bar 6}\psi^{S=0}_{(1\otimes1)}$ as well as $\psi^c_{\bar 3\otimes3}\psi^{S=0}_{(1\otimes1)}$ and $\psi^c_{6\otimes\bar 6}\psi^{S=0}_{(0\otimes0)}$.
There is no configuration mixing for $J=1$ and $J=2$ states since no cross term is found.

{Possible combinations of the color-spatial-spin-favor configurations of the $qq$ and $\bar q\bar q$ clusters, as shown in Table~\ref{tab:ssf}, lead to the total isospins for all the eigenstates in Table~\ref{tab:qparticle}.} The theoretical masses of the ground and first radial excited light tetraquarks of various quark configurations are listed in Table~\ref{tab:qparticle}, together with experimental data of some exotic mesons which will be reviewed and discussed separately in this section.

\begin{table*}[t]
\caption{\label{tab:qparticle} Ground and first radial excited light tetraquark masses, experimental data of some exotic mesons from the cited sources, and tentative assignments. $I$ denotes isospin.}
\begin{ruledtabular}
\begin{tabular}{ccccccccc}
J & $qq\bar q\bar q$ configurations & nS & $M^{cal}$(MeV) & Assignments & $M^{exp}$(MeV) &$\Gamma$(MeV) &  Process
\\
\hline
\multirow{10}*{$J=0$} &  \multirow{4}*{$|\psi^c_{\bar 3\otimes3}\psi^{S=0}_{(0\otimes0)}$,$\psi^c_{6\otimes\bar 6}\psi^{S=0}_{(1\otimes1)}\rangle$,} &  \multirow{2}*{1S} & 1431 & $f_0(1500)$ &$1473\pm5$&$108\pm9$ &  $p\bar p\to  (\eta \eta)\pi $~\cite{Uman:2006xb}
\\
&&& 1812 &... & ... & ... & ...
\\ \cline{3-4}
 &  \multirow{2}*{$I=0$} & \multirow{2}*{2S} & 1886 & ... & ... & ... & ...
\\
&&& 1986 &$f_0(2020)$  &  $2037\pm8$    &   $296\pm17$  &  $p\bar p\to  (\eta \eta)\pi $~\cite{Uman:2006xb}
\\ \cline{2-4}
& \multirow{5}*{$|\psi^c_{\bar 3\otimes3}\psi^{S=0}_{(1\otimes1)}$,$\psi^c_{6\otimes\bar 6}\psi^{S=0}_{(0\otimes0)}\rangle$,} & \multirow{3}*{1S} & \multirow{2}*{1676} &\multirow{2}*{$f_0(1710)$} &$1760\pm15^{+15}_{-10}$&$125\pm25^{+10}_{-15}$& $\psi(2s)\to\gamma\pi^+\pi^-(K^+K^-)$~\cite{Ablikim:2005kp}
\\
&&&&&$1759\pm6^{+14}_{-25}$&$172\pm10^{+32}_{-16}$ &  $J/\psi \to \gamma (\eta\eta)$~\cite{Ablikim:2013hq}
\\
& \multirow{3}*{$I=0,1,2$} && 2041 &$f_0(2020)$     &   $2037\pm8$    &   $296\pm17$  &  $p\bar p\to  (\eta \eta)\pi $~\cite{Uman:2006xb}
\\ \cline{3-4}
& & \multirow{2}*{2S} & 2141 & $f_0(2100)$ &$2081\pm13^{+24}_{-36}$&$273^{+27+70}_{-24-23}$ &  $J/\psi\to\gamma (\eta\eta)$~\cite{Ablikim:2013hq}
\\
& & & 2252 &$f_0(2200)$ &$2170\pm20^{+10}_{-15}$&$220\pm60^{+40}_{-45}$ &  $\psi(2s)\to\gamma\pi^+\pi^-(K^+K^-)$~\cite{Ablikim:2005kp}
\\
\hline
\multirow{10}*{$J=1$}& \multirow{2}*{$|\psi^c_{6\otimes\bar 6}\psi^{S=1}_{(1\otimes0)}\rangle$, $I=1$} & 1S & 1858 & $b_1(1960)$     &    $1960\pm35$    &    $230\pm50$   &  $p\bar p\to\omega \pi^0, \omega\eta\pi^0, \pi^+\pi^-$~\cite{Anisovich:2002su}
\\
& & 2S & 2262 & $b_1(2240)$     &    $2240\pm35$    &    $320\pm85$  &  $p\bar p\to\omega \pi^0, \omega\eta\pi^0, \pi^+\pi^-$~\cite{Anisovich:2002su}
\\\cline{2-4}
& \multirow{2}*{$|\psi^c_{\bar 3\otimes3}\psi^{S=1}_{(1\otimes0)}\rangle$, $I=1$} &1S & 1823 & ... & ... & ... & ...
\\
& &2S & 2280 & ... & ... & ... & ...
\\\cline{2-4}
& \multirow{2}*{$|\psi^c_{6\otimes\bar 6}\psi^{S=1}_{(1\otimes1)}\rangle$, $I=0$} &1S & 1678 & $h_1(1595)$     &    $1594\pm15^{+10}_{-60}$    &    $384\pm60^{+70}_{-100}$  &  $\pi^-p\to(\omega\eta)n$~\cite{Eugenio:2000rf}
\\
& &2S & 2081 & $h_1(1965)$     &    $1965\pm45$    &    $345\pm75$  &  $p\bar p\to\omega\eta, \omega\pi^0\pi^0$~\cite{Anisovich:2011sva}
\\\cline{2-4}
& \multirow{2}*{$|\psi^c_{\bar 3\otimes3}\psi^{S=1}_{(1\otimes1)}\rangle$, $I=0,1,2$}& 1S & 1875 & ... & ... & ... & ...
\\
& & 2S & 2331 & ... & ... & ... & ...
\\
\hline
\multirow{5}*{$J=2$}& \multirow{2}*{$|\psi^c_{6\otimes\bar 6}\psi^{S=2}_{(1\otimes1)}\rangle$, $I=0$}& 1S & 1936 &$X_2(1930)$ &$1930\pm25$&$450\pm50$ &  $\pi^-p\to(\eta\eta)n$~\cite{Binon:2004yd}
\\
& & 2S & 2339 & $f_2(2340)$ &$2362^{+31+140}_{-30-63}$&$334^{+62+165}_{-54-100}$ &  $J/\psi\to\gamma(\eta\eta)$~\cite{Ablikim:2013hq}
\\\cline{2-4}
& \multirow{2}*{$|\psi^c_{\bar 3\otimes3}\psi^{S=2}_{(1\otimes1)}\rangle$, $I=0,1,2$}& 1S & 1978 & $X_2(1980)$ &$1980\pm2\pm14$&$297\pm12\pm6$ &  $ \gamma\gamma\to (K^+K^-) $~\cite{Abe:2003vn}
\\
& & 2S & 2435 & $f_2(2300)$ &$2327\pm9\pm6$  &  $275\pm36\pm20$ &  $ \gamma\gamma\to (K^+ K^-) $~\cite{Abe:2003vn}
\\

\end{tabular}
\end{ruledtabular}
\end{table*}

\subsection{\label{sec:J0}$J=0$ States}

The meson mass spectrum has been studied by using the quark model for more than a half century. Especially, the heavy (c and b) flavor sector is well described by the NQM, and the predictions of NQM are accurate even for higher excited states. However, in the light meson region, the problem of understanding some exotic light mesons, firstly $f_0$ states, has puzzled people for many years.

For the states with $J^{PC}=0^{++}$, three isoscalar resonances: the $f_0(1370)$, $f_0(1500)$, and $f_0(1710)$ which are likely non-$q\bar q$ candidates are mainly reviewed in Ref.\cite{PDG}.  One conclusion reached is that none of the proposed $q\bar q$ ordering schemes in scalar multiplets is completely satisfactory. The $f_0(1370)$ and $f_0(1500)$ decay mostly into pions ($2\pi$ and $4\pi$), and the $f_0(1710)$ decays mainly into $K\bar K$ final states. Naively, one implies an $n\bar n(=u\bar u+d\bar d)$ structure for the $f_0(1370)$ and $f_0(1500)$, and an $s\bar s$ structure for the $f_0(1710)$.

However, the $1^3P_0$ state is always the lightest state in the three $1^3P_J$ states ($J=0,1,2$) in potential model studies~\cite{Godfrey:1985xj, Vijande:2004he, Ebert:2009ub, Xiao:2019qhl, Li:2020xzs}, which is confirmed in the observation of the $\chi_{cJ}(1P)$ and $\chi_{bJ}(1P)$ for charmonium and bottomonium mesons respectively. Since the mass splitting between $\chi_{c0}(1P)$ and $\chi_{c2}(1P)$ is around 150 MeV, and the mass splitting between $\chi_{b0}(1P)$ and $\chi_{b2}(1P)$ is around 50 MeV~\cite{PDG}, one may conclude that the $1^3P_0$ $n\bar n$ and $s\bar s$ states should be obviously lighter than $1^3P_2$ $n\bar n$ and $s\bar s$ states which are widely accepted as the $f_2(1270)$ and $f'_2(1525)$ respectively~\cite{Amsler:2004ps}. Thus, the $f_0(1500)$ and $f_0(1710)$ are too heavy to be accommodated as conventional mesons.

In $\gamma\gamma$ collisions, both of the $f_0(1500)$ and $f_0(1710)$ are not observed by ALEPH in $\gamma\gamma\to\pi^+\pi^-$ \cite{ALEPH:1999jnq}, and the $f_0(1500)$ is also not observed by Belle in $\gamma\gamma\to\pi^0\pi^0$ \cite{Belle:2008bmg}, which does not favor an $n\bar n$ interpretation for the $f_0(1500)$. Several glueball interpretations are proposed: the $f_0(1370)$ is mainly $n\bar n$, the $f_0(1500)$ mainly glueball, the $f_0(1710)$ dominantly $s\bar s$ \cite{Amsler:1995td, Close:2001ga}, or the $f_0(1710)$ as the glueball \cite{Janowski:2014ppa, Brunner:2015yha}.

The $f_0(1710)$ and $f_2(2200)$ are observed by Belle in $\gamma\gamma\to K^0_SK^0_S$ \cite{Belle:2013eck}. The mass, total width, and decay branching fraction to the $K\bar K$ state $\Gamma_{\gamma \gamma} B(K \bar K)$ are measured. One conclusion is that the $f_0(1710)$ and $f_2(2200)$ are unlikely to be glueballs because their total widths and $\Gamma_{\gamma \gamma} B(K \bar K)$ values are much larger than those expected for a pure glueball state. The $f_0(1500)$ is observed by BESII in $J/\psi \to \gamma\pi\pi$ \cite{Ablikim:2006db} and by BESIII in $J/\psi \to \gamma \eta \eta$ \cite{BESIII:2013qqz} with a much smaller rate than for the $f_0(1710)$, which speaks against a glueball interpretation of the $f_0(1500)$. Recently, The $f_0(1500)$ is studied in the framework of supersymmetric light front holographic QCD (LFHQCD) and identified as a isoscalar tetraquark \cite{Zou:2019tpo}.

As the review and discussion above, neither a conventional meson nor a glueball interpretation for the $f_0(1500)$ and $f_0(1710)$ is completely satisfactory.

The $f_0(1370)$ is assigned to be the $1^3P_0$ $s\bar s$ state by a recently quark model study of $s\bar s$ meson mass spectrum \cite{Li:2020xzs}, which is consistent with quark model mass spectrum studies \cite{Xiao:2019qhl, Vijande:2004he, Ebert:2009ub} but conflict with the experimental conclusion that the $f_0(1370)$ decays mostly into pions. Actually, since the average mass of the $f_0(1370)$ is from 1200 MeV to 1500 MeV \cite{PDG}, the broad $f_0(1370)$ resonance may correspond to two different states, each with the $n\bar n$ or $s\bar s$ content. Therefore, some resonances around 1370 MeV observed in the $K\bar K$ channel might be good candidates for the $1^3P_0$ $s\bar s$ state \cite{Li:2020xzs}.

Since both the $f_0(1500)$ and $f_0(2020)$ were observed by E835 in the process $p\bar p\to (\eta \eta)\pi $~\cite{Uman:2006xb}, we may group the $f_0(1500)$ and $f_0(2020)$ to be the ground states and first radial excited states respectively, with $J=0$, of the $|\psi^c_{\bar 3\otimes3}\psi^{S=0}_{(0\otimes0)}$,$\psi^c_{6\otimes\bar 6}\psi^{S=0}_{(1\otimes1)}\rangle$ mixed configuration.

Considering that both the $f_0(1710)$ and $f_0(2100)$ were observed by BESIII in the process $J/\psi \to \gamma (\eta\eta)$~\cite{Ablikim:2013hq}, and the $f_0(2020)$ was observed by E835 in the process $p\bar p\to (\eta \eta)\pi $ and their decay widths are in the same order \cite{Uman:2006xb}, and both the $f_0(1710)$ and $f_0(2200)$ were observed by BES in the process $\psi(2s)\to\gamma\pi^+\pi^-(K^+K^-)$ with the same order decay widths~\cite{Ablikim:2005kp}, we may assign the $f_0(1710)$ and $f_0(2020)$ to be the ground states, the $f_0(2100)$ and $f_0(2200)$ to be the first radial excited state with $J=0$ of the $|\psi^c_{\bar 3\otimes3}\psi^{S=0}_{(1\otimes1)}$,$\psi^c_{6\otimes\bar 6}\psi^{S=0}_{(0\otimes0)}\rangle$ mixed configuration, respectively.

Two states with masses $1812$ and $1886$ MeV are predicted in the calculation, which are close to $X(1835)$ and $X(1840)$. The $X(1835)$ is interpreted as a baryonium~\cite{Wang:2010vz, Deng:2012wi, Deng:2013aca} or the second radial excited state of $\eta'(958)$~\cite{Liu:2010tr, Yu:2011ta}. The $X(1835)$ has been observed and confirmed mainly by BESIII since 2005 \cite{BES:2005ega, BESIII:2011aa, BESIII:2013sbm, BESIII:2015xco, BESIII:2016fbr}, with the mass determined ranging from 1825 to 1910 MeV in various decay processes. The X(1840) is observed in the decay process $J/\psi\to\gamma 3(\pi^+\pi^-)$~\cite{BESIII:2013sbm}, and can theoretically take the $0^{++}$ quantum numbers. More experimental data in the $1800$ -- $1900$ MeV region are essential to reveal wether there might be more resonances in the mass region.

\subsection{\label{sec:J2}$J=2$ States}
Two well established $2^{++}$ states, the $f_2(1270)$ and $f'_2(1525)$, are widely accepted as the isoscalar $1^3P_2$ mesons for $n\bar n$ and $s\bar s$ structure respectively \cite{Amsler:2004ps}, which is consistent with the theoretical predictions of mesons \cite{Godfrey:1985xj, Vijande:2004he, Ebert:2009ub, Xiao:2019qhl, Li:2020xzs}. At higher masses, the $f_2(1950)$ and $f_2(2010)$ appear to be solid \cite{PDG}, and the $f_2(2010)$ is assigned to be the $2^3P_2$ $s\bar s$ state while $f_2(1950)$ does not fit into quark model spectrum easily \cite{Ebert:2009ub, Li:2020xzs}. Another two established tensor states, the $f_2(2300)$ and $f_2(2340)$, do not fit into quark model spectrum either.

The broad $f_2(1950)$ has been observed in several processes decaying to $4\pi$ \cite{WA102:1999lqn}, $\eta\eta$ \cite{Binon:2004yd}, and $K^+K^-$ \cite{Abe:2003vn}. Based on assuming that the $\eta\eta$ and the $K^+K^-$ are the dominant decay modes of the $f_2(1950)$, the $f_2(1950)$ is unlikely to be $n\bar n$ state. And it may not be a $s\bar s$ state too since the $2^3P_2$ $s\bar s$ state is occupied by the $f_2(2010)$~\cite{Li:2020xzs}. Meanwhile, the big mass difference of the two $f_2(1950)$ determined in the two processes $\pi^-p\to(\eta\eta)n$ \cite{Binon:2004yd} and $ \gamma\gamma\to (K^+K^-) $ \cite{Abe:2003vn} leads us to propose that they are likely two different states. We may use the $X_2(1930)$ and $X_2(1980)$ to represent the states of $\eta\eta$ and $K^+K^-$ decay modes respectively.

Since both the $X_2(1980)$ and $f_2(2300)$ are observed in the process $ \gamma\gamma\to (K^+K^-) $ with the similar decay widths \cite{Abe:2003vn}, one may naturally pair the $X_2(1980)$ and $f_0(2300)$ together. Therefore, we may assign the $X_2(1980)$ and $f_0(2300)$ to be the ground and first radial excited states, with $J=2$, of the $(\bar3_c \otimes 3_c)(1_s\otimes1_s)_{S=2}$ configuration, respectively.

Since both the $X_2(1930)$ and $f_2(2340)$ can decay to $\eta\eta$ and their decay widths are in the same order, we may group the $X_2(1930)$ and $f_2(2340)$ to be the ground and first radial excited tetraquark states respectively, with $J=2$, of the $(6_c \otimes \bar 6_c)(1_s\otimes1_s)_{S=2}$ configuration.

\subsection{\label{sec:J1}$J=1$ States}

With $J^{PC}=1^{+-}$, the $h_1(1170)$ and $h_1(1415)$ are convinced ground states of $n\bar n$ and $s\bar s$ isoscalar mesons respectively, and the $b_1(1235)$ is the ground state of isovector mesons in quark model \cite{Godfrey:1985xj, Li:2020xzs, PDG}.  However, the $h_1(1595)$ observed by BNL-E852 in the $\pi^-p\to(\omega\eta)n$ process \cite{Eugenio:2000rf}, the $h_1(1965)$ with a mainly decay channel $\omega \eta$ \cite{Anisovich:2011sva}, and the $b_1(1960)$ and $b_1(2240)$ observed in the process $p\bar p\to\omega \pi^0, \omega\eta\pi^0, \pi^+\pi^-$ \cite{Anisovich:2002su} do not fit into the $q\bar q$ meson mass spectrum.

The main decay channel of the $h_1(1595)$ and $h_1(1965)$, $\omega \eta$, is observed for neither the $h_1(1170)$ nor $h_1(1415)$ while the decay widths of the $h_1(1595)$ and $h_1(1965)$ are in the same order, one may tentatively pair the $h_1(1595)$ and $h_1(1965)$ together and separate them from conventional mesons. We may group the $h_1(1595)$ and $h_1(1965)$ to be the ground state and first radial excited states respectively, with $J=1$, of the $(6_c \otimes \bar6_c)(1_s\otimes1_s)_{S=1}$ configuration.

We may tentatively assign the $b_1(1960)$ and $b_1(2240)$ to be the ground and first radial excited states, with $J=1$, of the $(6_c \otimes \bar 6_c)(1_s\otimes0_s)_{S=1}$ configuration, respectively. The $b_1(1960)$ and $b_1(2240)$ are paired since they are observed in the process $p\bar p\to\omega \pi^0, \omega\eta\pi^0, \pi^+\pi^-$ \cite{Anisovich:2002su} and their decay widths are in the same order.  There are very rare experimental data for $b_1$ states except for the established $b_1(1235)$, and the $b_1(1960)$ and $b_1(2240)$ are not established states in PDG~\cite{PDG}. More experimental data for $b_1$ states are required to make more unambiguous assignments.

As shown in Table \ref{tab:qparticle}, the ground and first radial excited $J=0,1,2$ light tetraquark states predicted in the work have been tentatively matched with experimental data in pairs. We have provided in the work a possible tetraquark interpretation for some exotic meson states. 
For the interpretation that those exotic particles might be the mixture of $q\bar q$ meson, glueball, and tetraquark, one may refer to Refs~\cite{Klempt:2021ope,Klempt:2021nuf,Sarantsev:2021ein}.
{It may be suggested that pure tetraquark states are searched experimentally via double-charged channels since $I=2$ tetraquarks are predicted.}

\section{Summary}\label{sec:SUM}

The masses of ground and first radial excited light tetraquark states have been evaluated, with all model parameters predetermined by fitting the masses of light, charmed and bottom mesons. A tentative matching has been made between the predicted ground and first radial excited light tetraquark states and the believed exotic mesons.

For $J=0$ states, the work suggests that the $f_0(1500)$ and $f_0(2020)$ might be the ground and first radial excited states respectively of the $|\psi^c_{\bar 3\otimes3}\psi^{S=0}_{(0\otimes0)}$,$\psi^c_{6\otimes\bar 6}\psi^{S=0}_{(1\otimes1)}\rangle$ mixed configuration, and that the $f_0(1710)$ and $f_0(2020)$ might be the ground states, and the $f_0(2100)$ and $f_0(2200)$ might be the first radial excited states of the $|\psi^c_{\bar 3\otimes3}\psi^{S=0}_{(1\otimes1)}$,$\psi^c_{6\otimes\bar 6}\psi^{S=0}_{(0\otimes0)}\rangle$ mixed configuration, respectively.

For $J=2$ states, we first assume that the $f_2(1950)$ may represent two different resonances because of the large mass difference of the $f_2(1950)$ determined in the two processes $\pi^-p\to(\eta\eta)n$ \cite{Binon:2004yd} and $ \gamma\gamma\to (K^+K^-) $ \cite{Abe:2003vn}. Then we have tentatively assigned the $X_2(1980)$ and $f_0(2300)$ to be the ground and first radial excited states of the $(\bar3_c \otimes 3_c)(1_s\otimes1_s)_{S=2}$ configuration, respectively, and the $X_2(1930)$ and $f_2(2340)$ to be the ground and first radial excited tetraquark states respectively of the $(6_c \otimes \bar 6_c)(1_s\otimes1_s)_{S=2}$ configuration.

For $J=1$ states, the work supports that the $h_1(1595)$ might be  the ground light tetraquark state of the $(6_c \otimes \bar6_c)(1_s\otimes1_s)_{S=1}$ configuration, and the $h_1(1965)$ might be the first radial excited state of the $h_1(1595)$. The assignment of the $b_1(1960)$ and $b_1(2240)$ is rather ambiguous in the work, that is, the $b_1(1960)$ and $b_1(2240)$ may be paired to be the ground and first radial excited states respectively of the $(6_c \otimes \bar 6_c)(1_s\otimes0_s)_{S=1}$ configuration.

\begin{acknowledgments}
This work was supported by (i) Suranaree University of Technology (SUT), (ii) Thailand Science Research and Innovation (TSRI), and (iii) National Science Research and Innovation Fund (NSRF), project no. 160355.  X.Y. Liu acknowledges support from the Young Science Foundation from the Education Department of Liaoning Province, China (Project No. LQ2019009).
\end{acknowledgments}

\bibliography{PRD2021}

\end{document}